\documentclass[a4paper]{spie}  

\usepackage{amsmath,amsfonts,amssymb,enumitem}
\usepackage{graphicx}
\usepackage{graphbox}
\usepackage[colorlinks=true, allcolors=blue]{hyperref}
\usepackage[T1]{fontenc}
\usepackage[utf8]{inputenc}
\usepackage{babel}
\usepackage[font=small,labelfont=bf]{caption}
\usepackage[capitalise]{cleveref}
\usepackage{wrapfig}
\usepackage{enumitem}
\usepackage{array}
\newcolumntype{P}[1]{>{\centering\arraybackslash}p{#1}}

\newcommand{\vx}{\mathbf{x}}
\newcommand{\vxhat}{\widehat{\mathbf{x}}}
\newcommand{\vxprime}{\mathbf{x}^{\prime}}

\newcommand{\vz}{\mathbf{z}}

\newcommand{\vr}{\mathbf{r}}

\newcommand{\mux}{\boldsymbol{\mu}({\mathbf{x}})}
\newcommand{\sigmax}{\boldsymbol{\sigma}(\mathbf{x})}

\newcommand{\normal}[2]{\mathcal{N}\left(#1, #2\right)}

\title{Leveraging healthy population variability in deep learning unsupervised anomaly detection in brain FDG PET}

\author[a]{Ma\"{e}lys Solal}
\author[a]{Ravi Hassanaly}
\author[a]{Ninon Burgos}
\affil[a]{Sorbonne Université, Institut du Cerveau - Paris Brain Institute - ICM, CNRS, Inria, Inserm, AP-HP, Hôpital de la Pitié Salpêtrière, F-75013, Paris, France}

\authorinfo{Further author information: (Send correspondence to Ma\"{e}lys Solal)\\Ma\"{e}lys Solal: E-mail: maelys.solal@icm-institute.org\\
Ninon Burgos: E-mail: ninon.burgos@cnrs.fr}

\begin{document}
\maketitle

\begin{abstract}
Unsupervised anomaly detection is a popular approach for the analysis of neuroimaging data as it allows to identify a wide variety of anomalies from unlabelled data. It relies on building a subject-specific model of healthy appearance to which a subject's image can be compared to detect anomalies. In the literature, it is common for anomaly detection to rely on analysing the residual image between the subject's image and its pseudo-healthy reconstruction. This approach however has limitations partly due to the pseudo-healthy reconstructions being imperfect and to the lack of natural thresholding mechanism. Our proposed method, inspired by Z-scores, leverages the healthy population variability to overcome these limitations. Our experiments conducted on FDG PET scans from the ADNI database demonstrate the effectiveness of our approach in accurately identifying Alzheimer's disease related anomalies.
\end{abstract}


\section{Introduction}
\label{sec:intro}  

Alzheimer’s disease (AD) is a prevalent neurodegenerative disorder characterised by progressive memory loss and cognitive decline. Early detection of AD-related brain abnormalities is crucial, and neuroimaging such as \textsuperscript{18}F-fluorodeoxyglucose (FDG) positron emission tomography (PET) has the potential to play a vital role in this process, by allowing to visualise subtle changes in neuronal activity before the onset of clinical symptoms\cite{mosconi2005brain}.

Machine learning (ML) methods have shown promise in the quantitative analysis of neuroimaging data, which is key in detecting subtle changes in brain patterns amidst natural variations. This work aligns with the unsupervised anomaly detection (UAD) framework, where ML algorithms are used to identify rare events from unlabelled data \cite{chen2022unsupervised}. The use of UAD algorithms is motivated by the limited availability of labelled data, and the hope that these methods enable the detection of various types of anomalies, including those associated with AD. 

This approach relies on generating a subject-specific model of healthy appearance, and comparing the subject's real image with the model, resulting in a subject-specific map of anomalies. In the training phase, the neural network is exclusively shown images of healthy subjects, enabling it to learn the distribution of healthy images. When applied to a patient's image (with an unknown diagnosis) during the testing phase, the network reconstructs the image based on its knowledge of healthy images \cite{chen2022unsupervised}. Consequently, we hypothesise that the reconstructed image is a subject-specific pseudo-healthy representation of the input image. By comparing the input image with its reconstruction, areas of the brain exhibiting anomalies can be effectively highlighted. 

One popular method in anomaly detection involves analysing the residual error between a patient's image and its pseudo-healthy reconstruction \cite{baur_autoencoders_2021,chen2022unsupervised}. This approach however has several shortcomings \cite{meissen2021pitfalls}. Firstly, there is no clear thresholding mechanism to selectively focus on significant anomalies, which leads to a higher number of false positives. Secondly, the method does not consider the normal variability within the dataset, making it harder to capture subtle anomalies. Additionally, it fails to account for the reconstruction error of the model itself, which may be evident in images from healthy individuals. 

Our proposed approach aims to address the aforementioned limitations by introducing abnormality maps constructed by relying loosely on Z-scores. Our method leverages variability within the training dataset to enhance anomaly detection. It also provides improved interpretability, as it allows for a direct association between the abnormality map and the standard deviation within a control population, thereby facilitating the identification of significant anomalies by establishing a clear threshold. Consequently, our approach offers a more comprehensive solution to design abnormality maps that better account for model errors, and are more interpretable, by leveraging the intuitive nature of Z-scores.

\section{Materials and Methods}

\subsection{Dataset}

In this study, we work with FDG PET scans from the ADNI database \cite{mueller2005alzheimer, Jagust2010AlzheimerDisease, Jagust2015AlzheimerDisease}. We select images that are co-registered, averaged and uniformised to the same resolution. These images are then further pre-processed using the \texttt{pet-linear} pipeline from Clinica \cite{clinica} which registers to a standard space, normalises in intensity and crops each image. For our purposes, we work with images from healthy subjects, labelled as cognitively normal (CN), and some images of diseased subjects, labelled as Alzheimer's disease (AD). In total, we include 247 CN subjects (which amounts to 521 images) and 273 AD subjects, with only one image each. 

\subsection{Generative model - 3D VAE}

\begin{wrapfigure}{l}{0.39\textwidth}
    \centering
    \includegraphics[width=\linewidth]{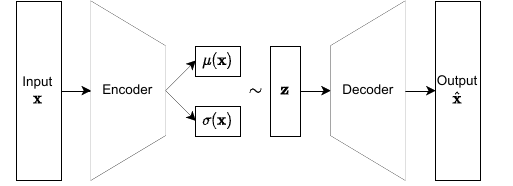}
    \caption{Variational autoencoder (VAE) architecture}
    \label{fig:vae}
\end{wrapfigure}

To generate subject-specific models of healthy appearance, we rely on variational autoencoders (VAEs) \cite{kingma2014autoencoding}, a type of deep probabilistic generative model. During the training phase, the model is shown only images labelled as cognitively normal (CN) to learn the distribution of healthy images. At inference time, the image of a subject is given as input to the previously trained VAE, which generates an image that looks healthy and corresponds to that subject specifically. By comparing the input image and its pseudo-healthy reconstruction outputted by the VAE, we can detect and localise the areas of the brain which present anomalies.

A VAE aims to approximate the true distribution of the training data with a simple parameterised distribution (as shown schematically in \cref{fig:vae}). In practice, the encoder projects an input sample $\vx$ to a latent space of smaller dimension which is constrained to parameterise a standard normal distribution. Each input is thus mapped to a normal distribution $\normal{\mux}{\sigmax^2\mathbf{I}}$. The latent vector $\vz$ is then sampled from this distribution, and the decoder learns to reconstruct $\vxhat$ from $\vz$. 

\subsection{Abnormality maps} 

Once the subject-specific pseudo-healthy image $\vxhat$ is generated, it needs to be compared to the input image $\vx$ to generate an abnormality map from which anomalies can be detected. In the literature, it is common to define the abnormality map as the voxel-wise difference between these two images $\vr = \vx - \vxhat$, also known as the residual image. Unfortunately, this method does not account for the normal variability in the population and the potential reconstruction errors of our generative model. In other words, the residual errors are considered at the same level of importance for every voxel, even in regions which are highly variable between subjects.

In a typical voxel-wise analysis \cite{minoshima1995diagnostic,drzezga2005prediction}, a subject's PET image is registered to a standard space and compared voxel-by-voxel to a distribution obtained from normal control scans. This can be done through a Z-score, or standard score, given by the ratio of the difference between an individual raw value and the control population mean, and the control population standard deviation. The score then corresponds to the number of standard deviations by which the observed value deviates from the mean value in a control population. 
Classically, this method also allows setting a threshold, corresponding to a level of confidence. 

In our work, we generate abnormality maps which rely loosely on the idea of Z-scores since this statistic is widely used among clinicians. We set:
$
\tilde{\vz} = \frac{\vx - \vxhat}{\sigma},
$
where $\vx$ is the input image, $\vxhat$ is its pseudo-healthy reconstruction, and $\sigma$ is the standard deviation of the set of healthy images used to train our model (in some way, our healthy control population).

\subsection{AD hypometabolism simulation framework}

Evaluating our ability to detect anomalies is challenging due to the lack of ground truth. Indeed, we only have access to unlabelled images, with a diagnosis label but no anomaly annotations, that is, neither contour nor mask. We can however rely on the hypometabolism simulation framework proposed by Ref.~\citenum{hassanaly2023simulation}. It consists of a new test set, where hypometabolism characteristic of Alzheimer's disease (or other dementia subtypes) can be simulated on healthy images to have pairs of healthy and diseased images with known anomalies. For a given healthy image $\vx$, we obtain a corresponding image $\vxprime$ with a certain degree of hypometabolism (in our case, 30\%). To validate our ability to detect anomalies, we can compare our abnormality map (or residual) with the mask that was used to simulate the hypometabolism between $\vx$ and $\vxprime$ in the first place.

\subsection{Experimental setup}

To avoid data leakage, we split our dataset into distinct splits. The images of the AD subjects are used solely during testing. We perform training, validation and testing data splits of the CN dataset at the subject's level and stratified by sex and age to reduce biases. The test set includes 50 images from 50 CN subjects, the validation set includes 19 images from 19 subjects, and the training set includes 452 images from 178 subjects. 

To synthesise images of healthy appearance, we rely on a 3D VAE proposed in Ref.~\citenum{hassanaly2023simulation}. Its encoder is composed of three convolutional layers and one dense layer, and its decoder is symmetrical. After each convolutional layer is a batch normalisation and a leaky-ReLU activation function. The latent space was chosen to have a size of 256. The model was trained on 300 epochs, with a learning rate of $10^{-5}$ and a batch size of 8 using ClinicaDL \cite{thibeau-sutreClinicaDLOpensourceDeep2022}. 

\section{Results}

\begin{figure}[h!]
    \includegraphics[width=\linewidth]{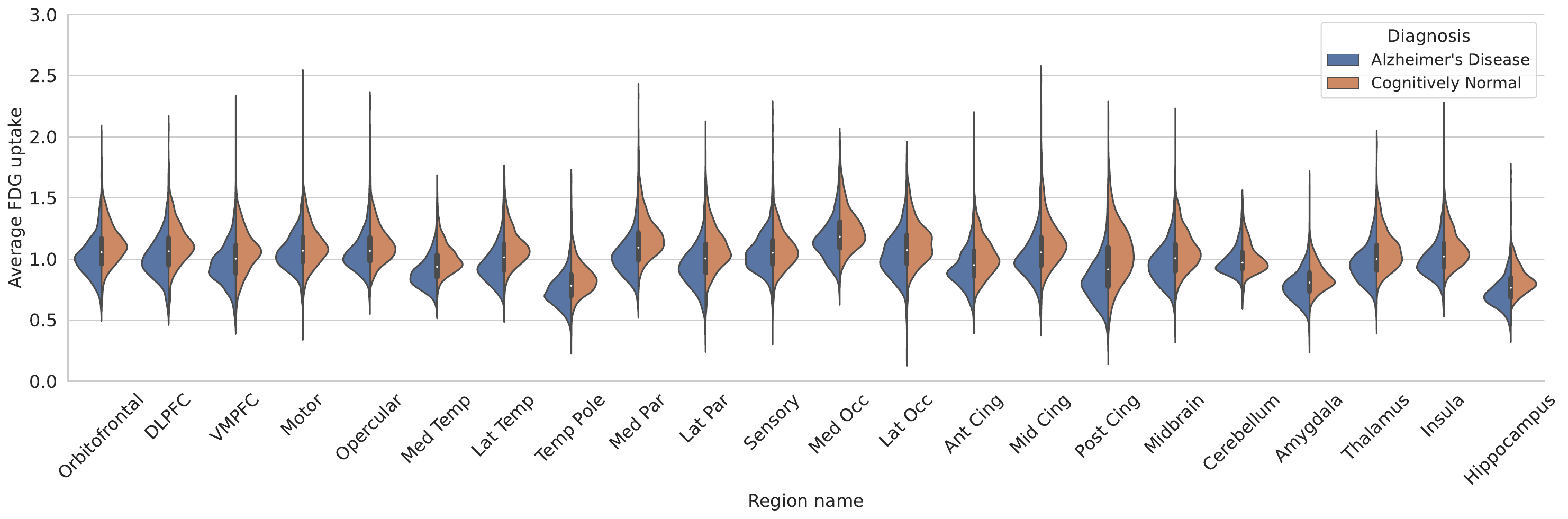}
    \caption{Preliminary analysis: comparison of the distribution of average regional FDG uptake in cognitively normal (orange) and Alzheimer's disease (blue) subjects}
    \label{fig:prelim}
\end{figure}

To assess the inter- and intra-subject variability of our data, we performed both a voxel-wise and a regional statistical analysis, and report the results of the latter in \cref{fig:prelim}. Both analyses confirms our intuition that some regions exhibit a larger variability than others, making the voxels in that region more difficult to reconstruct, and therefore more prone to errors when performing anomaly detection. It therefore seems coherent to attempt to account for this variability in our abnormality maps. For instance, the threshold to consider a voxel as abnormal in regions that exhibit high inter-subject variability could be set higher than in those with lower variability. 

\begin{figure}[htbp]
    \centering
    \includegraphics[width=\linewidth]{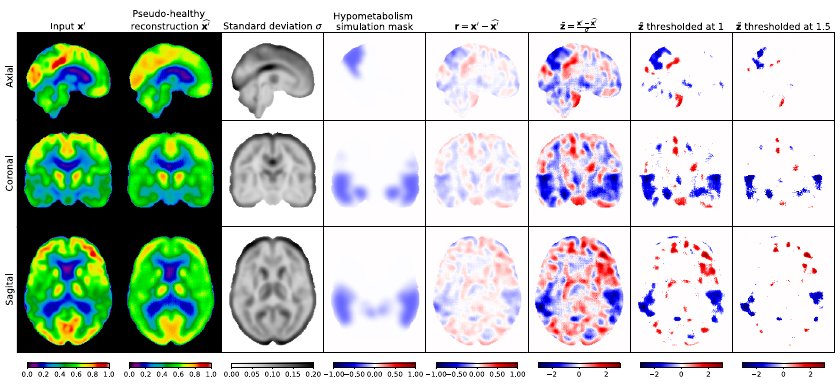}
    \caption{Example of results obtained for a healthy image with simulated hypometabolism. We display a central slice for each plane with the input $\vx$, the pseudo-healthy reconstruction $\vxhat$, the standard deviation of the training set of CN images $\sigma$, the hypometabolism simulation mask with the anomalies we aim to retrieve, the residual $\vr = \vx - \vxhat$, the abnormality map $\tilde{\vz} = \frac{\vx - \vxhat}{\sigma}$, and the abnormality map $\tilde{\vz}$ thresholded at 1 and 1.5.}
    \label{fig:abn_maps}
\end{figure}

We plot the results of residual-based and Z-score-based abnormality maps for a subject with simulated hypometabolism in \cref{fig:abn_maps}. We find that Z-score-based abnormality maps indeed better highlight relevant anomalies simulated with the hypometabolism mask through the use of thresholds, thus reducing the impact of the model's reconstruction error.

To complement our visual analysis of the abnormality maps, we compute the normalised cross-correlation coefficient (NCC) between the hypometabolism simulation mask and our abnormality maps. The average NCC across the testing set between the mask and the residual $\vr$ is 0.441 ($\pm 0.115$), and between the mask and our new abnormality map $\tilde{\vz}$ is 0.487 ($\pm 0.117$). We find that the Z-score-based abnormality maps allow us to better retrieve the anomalies that were simulated, as shown by a larger correlation magnitude.

\section{Discussion \& Conclusion}

Our work explores the detection of AD-related brain abnormalities using FDG PET neuroimaging. We adopt a UAD approach which relies on building subject-specific pseudo-healthy neuroimages to which a patient's image can be compared, as it provides the ability to detect subtle changes without prior knowledge. Traditional methods rely on residual analysis which has limitations which include the lack of natural thresholding mechanism, and its failure to consider normal variability and reconstruction errors\cite{baur_autoencoders_2021,chen2022unsupervised,meissen2021pitfalls}.

To address these limitations, we propose a novel approach that leverages healthy population variability to construct abnormality maps. Our method is inspired from Z-scores and offers improved interpretability by establishing a clearer association between the anomalies and the standard deviation in a control population. The integration of Z-scores also facilitates the identification of significant anomalies by providing a more natural thresholding mechanism. Our experimental results on FDG PET scans from the ADNI database highlight the potential of our method in accurately identifying brain abnormalities related to AD. Future works could involve validating this approach with other types of anomalies, such as those linked with other types of dementia.

Whilst our method allows to build more interpretable abnormality maps that highlight significant anomalies, it remains challenging to distinguish reconstruction errors caused by the model being imperfect from those related to anomalies in the input. Future efforts could be directed towards improving the reconstruction capabilities of the generative model, especially since VAEs are known for producing blurry reconstructions \cite{chen2022unsupervised}, with diffusion models \cite{pandey2022diffusevae} or conditional GANs for instance \cite{yaakub2019pseudo}. We could also try post-processing techniques to mitigate the effects of reconstruction errors, or generating abnormality maps using larger regions instead of voxel-wise analysis.

In summary, our proposed approach combines the advantages of UAD, such as the ability to detect subtle changes without prior knowledge, with the interpretability offered by Z-scores. Our experiments provide promising results, indicating that our method holds the potential for more reliable early detection of AD-related brain abnormalities.

\acknowledgments 
The research leading to these results has received funding from the French government under management of Agence Nationale de la Recherche as part of the "Investissements d'avenir" program, reference ANR-19-P3IA-0001 (PRAIRIE 3IA Institute) and reference ANR-10-IAIHU-06 (Agence Nationale de la Recherche-10-IA Institut Hospitalo-Universitaire-6).

The ADNI was launched in 2003 as a public-private partnership, led by Principal Investigator Michael W. Weiner, MD. The primary goal of ADNI has been to test whether serial MRI, PET, other biological markers, and clinical and neuropsychological assessment can be combined to measure the progression of mild cognitive impairment and early AD.

\bibliography{ref} 
\bibliographystyle{spiebib} 

\end{document}